\documentclass[twocolumn,aps,unsortedaddress,showpacs,amsmath]{revtex4}

\usepackage{graphics,graphicx}

\begin{document}

\title{Probability density function of turbulent velocity fluctuations \\ in rough-wall boundary layer}
\thanks{To appear in Physical Review E}

\author{Hideaki Mouri}
\email{hmouri@mri-jma.go.jp}
\affiliation{Meteorological Research Institute, Nagamine 1-1, Tsukuba 305-0052, Japan}

\author{Masanori Takaoka}
\email{mtakaoka@mail.doshisha.ac.jp}
\affiliation{Department of Mechanical Engineering, Doshisha University, Kyotanabe, Kyoto 610-0321, Japan}

\author{Akihiro Hori}
\altaffiliation[]{Also at Meteorological and Environmental Service, Inc., Tama, Tokyo 206-0012, Japan}

\author{Yoshihide Kawashima}
\altaffiliation[]{Also at Meteorological and Environmental Service, Inc., Tama, Tokyo 206-0012, Japan}
\affiliation{Meteorological Research Institute, Nagamine 1-1, Tsukuba 305-0052, Japan}


\begin{abstract}

The probability density function of single-point velocity fluctuations in turbulence is studied systematically using Fourier coefficients in the energy-containing range. In ideal turbulence where energy-containing motions are random and independent, the Fourier coefficients tend to Gaussian and independent of each other. Velocity fluctuations accordingly tend to Gaussian. However, if energy-containing motions are intermittent or contaminated with bounded-amplitude motions such as wavy wakes, the Fourier coefficients tend to non-Gaussian and dependent of each other. Velocity fluctuations accordingly tend to non-Gaussian. These situations are found in our experiment of a rough-wall boundary layer.

\end{abstract}

\pacs{47.27.Ak, 47.27.Nz}

\maketitle

\section{INTRODUCTION}
\label{S1}

Suppose that single-point velocity fluctuations $u(x)$ are measured repeatedly in stationary turbulence over the range $0 \le x < L$. If the sampling interval is much greater than the eddy turnover timescale, these measurements serve as independent realizations of the turbulence. Each of them is expanded into a Fourier series as
\begin{equation}
\label{eq1}
u(x) = \sqrt{\frac{2}{L}} 
       \sum_{n=1}^{\infty} a_n \cos \left( \frac{2 \pi nx}{L} \right)
     +                     b_n \sin \left( \frac{2 \pi nx}{L} \right),
\end{equation}
where $2 \pi n/L = k_n$ is the wave number. Batchelor \cite{B53} assumed that the Fourier coefficients $a_n$ and $b_n$ are statistically independent of each other and applied the central limit theorem to their sum. This theorem ensures that the probability density function (PDF) of a sum of many independent random variables tends to Gaussian, at least within a few standard deviations around the average \cite{F71,KS77}. It was concluded that  velocity fluctuations tend to Gaussian, being consistent with experimental and observational data of turbulence that were available at that time.

However, recent measurements revealed the presence of velocity fluctuations that tend to non-Gaussian. Sreenivasan and Dhruva \cite{SD98} obtained long data of atmospheric turbulence at 35 m above the ground. Their data yield the flatness factor $F_u = \langle u^4 \rangle / \langle u^2 \rangle ^2 = 2.66$, where the bracket $\langle \cdot \rangle$ denotes an average. This value is different from the Gaussian value $F_u = 3$. Noullez et al. \cite{NWLMF97} obtained $F_u \simeq 2.85$ in turbulent jets. For these results, there has been no explanation.

Therefore, although single-point velocity fluctuations are fundamental in describing turbulence, the mechanism that determines their PDF is uncertain. We systematically study velocity fluctuations in a laboratory rough-wall boundary layer, a representative turbulent flow with various applications, e.g., the atmosphere near the ground. With an increase of the distance from the wall, velocity fluctuations are found to change from sub-Gaussian ($F_u < 3$) to Gaussian, and to hyper-Gaussian ($F_u > 3$). This behavior is discussed using PDFs of the Fourier coefficients and correlations among them.

\section{CONDITION FOR GSAUSSIANITY}
\label{S2}

This section serves as a summary of conditions for velocity fluctuations to be approximately Gaussian. It is assumed that the turbulence is not only stationary but also homogeneous in the $x$ direction. The data length is set to be much larger than the correlation length $l_c = \int \vert \langle u(x+\delta x)u(x) \rangle \vert d\delta x / \langle u^2 \rangle$. Then an average taken over the realizations is equal to the corresponding average taken over the $x$ positions.

Velocity fluctuations of turbulence are dominated by Fourier coefficients in the energy-containing range. Since the data length is large, there is a sufficient number of Fourier coefficients for the central limit theorem to be applicable. Velocity fluctuations tend to Gaussian if the Fourier coefficients are random and independent. This is expected for ideal turbulence where energy-containing motions are random and independent. Although turbulence always contains small-scale coherent structures, e.g., vortex tubes \cite{SA97}, their contribution to velocity fluctuations is as small as the energy ratio of the dissipation range to the energy-containing range.

The central limit theorem is not applicable to a sum of variables if few of them dominate over the others \cite{F71,KS77}. For example, if the energy spectrum $E_n = \langle a_n^2 \rangle + \langle b_n^2 \rangle$ is proportional to $k_n^{\alpha}$ with $\alpha < -1$, Fourier coefficients at the smallest wave numbers dominate the velocity fluctuations. They do not necessarily tend to Gaussian \cite{J98}. Nevertheless, the energy spectrum is relatively flat in the usual energy-containing range, where the power law $k_n^{\alpha}$ is not a good approximation. We previously assumed that the central limit theorem is not applicable to the sum of Fourier coefficients in any turbulence \cite{M02}, but this assumption was wrong.

Fourier coefficients of velocity fluctuations also tend to Gaussian if turbulence is made of random and independent motions \cite{M02,BP01,FB95}. The Fourier coefficient $a_n$ is obtained as 
\begin{eqnarray}
\label{eq2}
a_n 
&=& \sqrt{\frac{2}{L}} \int_{0}^{L} u(x) \cos \left( \frac{2 \pi n x}{L} \right) dx 
\nonumber \\
&=& \sqrt{\frac{2}{L}} \left( \int_{0}^{L/m}      ... \, dx +
                              \int_{L/m}^{2L/m}   ... \, dx 
\right. \nonumber \\ & & \left.  \qquad \qquad  + ... +
                              \int_{(m-1)L/m}^{L} ... \, dx
                       \right).
\end{eqnarray}
For $n \gg 1$, we are able to set $1 \ll m \ll n$ and $l_c \ll L/m$. The integrations $\int_{0}^{L/m} ...\, dx$, ..., and $\int_{(m-1)L/m}^{L} ... \, dx$ are regarded as random variables. Their amplitudes are the same because the segment size $L/m$ exceeds the wavelength $L/n$. They are nearly independent because adjacent integrations depend on each other only at the ends. Then $a_n$ tends to Gaussian as a consequence of the central limit theorem. This discussion is not applicable to Fourier coefficients $a_n$ and $b_n$ for $n \simeq 1$. They are nevertheless small and do not contribute to velocity fluctuations if the data length $L$ is sufficiently large, because of the universal trend $E_n \rightarrow 0$ in the limit $k_n = 2\pi n/L \rightarrow 0$ \cite{B53}. We are able to assume safely that all the Fourier coefficients are Gaussian.

The central limit theorem offers no information about the tails of the PDF of a sum of variables. Only when the sum (minus its mean) has been divided by the square root of the number of the variables, the Gaussian approximation holds at the tails \cite{F71,KS77}. Although statistics such as the flatness factor mainly reflect the core of the PDF, the tails that are significantly far from Gaussian could exist and affect the statistics. One example is significant contamination with a bounded-amplitude motion, e.g., $u(x) \propto \sin (x)$, the PDF of which does not have tails. Velocity fluctuations tend to sub-Gaussian, e.g., $F_u = 3/2$ for $u(x) \propto \sin (x)$ \cite{J98}. The Fourier coefficients that correspond to the bounded-amplitude motion also tend to sub-Gaussian \cite{footnote2}. Another example is significant intermittency, where velocity fluctuations tend to hyper-Gaussian \cite{SG00}. The Fourier coefficients also tend to hyper-Gaussian because some of the integrations $\int_{0}^{L/m} ...\, dx$, ..., and $\int_{(m-1)L/m}^{L} ... \, dx$ are enhanced over the others. Since the Fourier coefficients are linear functions of velocity fluctuations, sub- and hyper-Gaussianities of the velocity fluctuations yield sub- and hyper-Gaussianities of the Fourier coefficients, respectively, and vice versa. We expect that this relation could approximately exist even in the case of mutually dependent Fourier coefficients as long as the dependence is weak.

\begin{figure}[!]
\resizebox{9cm}{!}{\includegraphics*[0cm,13cm][20cm,27cm]{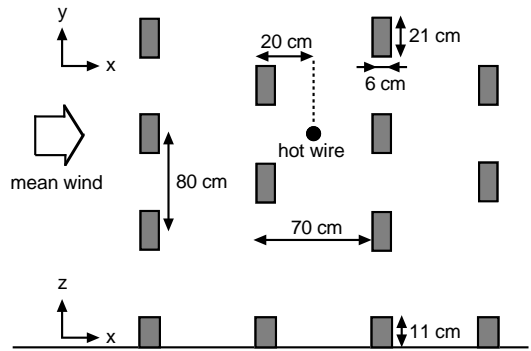}}
\caption{\label{F1} Schematic representation of our experimental setup. Plan and side views are shown together with coordinate axes (upper and lower sketches, respectively).}
\end{figure}

\begin{figure}[!]
\resizebox{8cm}{!}{\includegraphics*[4cm,5.5cm][19cm,27cm]{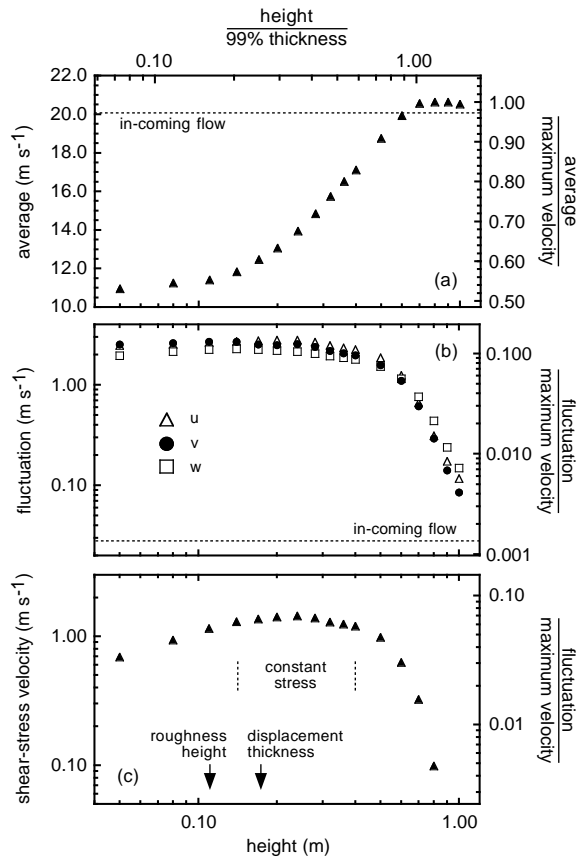}}
\caption{\label{F2} (a) Mean streamwise velocity $U$. (b) Root-mean-square velocity fluctuations $\langle u^2 \rangle ^{1/2}$, $\langle v^2 \rangle ^{1/2}$, and $\langle w^2 \rangle ^{1/2}$. (c) Shear-stress velocity $\langle -uw \rangle ^{1/2}$. The abscissa is the height $z$. The triangles denote the $u$ component, the circles denote the $v$ component, and the squares denote the $w$ component. The horizontal dotted lines indicate the values of $U$ and $\langle v^2 \rangle ^{1/2}$ obtained in the in-coming flow. The arrows indicate the height of the roughness elements and the displacement thickness. On the uppermost axis, we indicate the height normalized by the 99\% thickness. On the right axis, we indicate the values normalized by the maximum $U$ value. We also indicate the $z$ range of the constant-stress sublayer. The shear-stress velocity is not available at $z = 0.90$ and 1.00 m, where the correlation $-uw$ is negative.}
\end{figure}

\section{EXPERIMENT}
\label{S3}

The experiment was done in a wind tunnel of the Meteorological Research Institute. As shown in Fig. 1, we use the coordinates $x$, $y$, and $z$ in the streamwise, spanwise, and floor-normal directions. The corresponding wind velocities are $u$, $v$, and $w$. We take the origin $x=y=z=0$ on the tunnel floor at the entrance to the test section. The test section had the size $\delta x = 18$ m, $\delta y = 3$ m, and $\delta z = 2$ m. A boundary layer was made by placing blocks over the entire floor of the test section. The blocks had the size $\delta x = 6$ cm, $\delta y = 21$ cm, and $\delta z = 11$ cm. The spacings of the adjacent blocks were $\delta x = 70$ cm and $\delta y = 80$ cm. We set the in-coming wind velocity to be 20 m~s$^{-1}$. The boundary layer was well developed at $x \gtrsim 10$ m.

The $u$ and $v$ or $u$ and $w$ components of the wind velocity were measured using a hot-wire anemometer with an X-type probe. The wires were made of tungsten, 5 $\mu$m in diameter, 1.0 mm in effective length, 1.4 mm in separation, oriented at $\pm 45^{\circ}$ to the streamwise direction, and operated at the temperature of $280^{\circ}$C. The measurement positions were at $x = 10$ m and $z = 0.05$--1.00 m. The signal was low-pass filtered with 24 dB/octave and sampled digitally with 16-bit resolution. At $z < 0.30$ m, the filtering frequency was 10 kHz and the sampling frequency was 20 kHz. At $z > 0.30$ m, they were 25 kHz and 50 kHz, respectively. The length of the signal was $8 \times 10^6$ points at $z = 0.11$, 0.14, 0.24, 0.28, 0.40, 0.50, 0.60, 0.80, and 1.00 m or $32 \times 10^6$ points at $z = 0.05$, 0.08, 0.17, 0.20, 0.32, 0.36, 0.70, and 0.90 m. We used the frozen-eddy hypothesis of Taylor to convert temporal variations into spatial variations.

Figure~\ref{F2} shows the mean streamwise velocity $U$, the root-mean-square velocity fluctuations $\langle u^2 \rangle^{1/2}$, $\langle v^2 \rangle^{1/2}$, and $\langle w^2 \rangle^{1/2}$, and the shear-stress velocity $\langle -uw \rangle^{1/2}$ as a function of the height $z$. The 99\% thickness, i.e., the height at which $U$ is 99\% of its maximum value $\hat{U}$, is 0.68 m. The displacement thickness $\textstyle \int_0^{\infty} (1-U/\hat{U}) dz$ is 0.17 m \cite{footnote3}. For reference, the average $U$ and the root-mean-square fluctuation $\langle v^2 \rangle^{1/2}$ obtained in the in-coming flow at $x= z= 1$ m are shown. The latter reflects the mechanical and electric noise, which is well below the turbulence signals.

Throughout the boundary layer, velocity fluctuations are almost isotropic [Fig.~\ref{F2}(b)]. Although velocity fluctuations in a smooth-wall boundary layer are anisotropic, the anisotropy is reduced over roughness \cite{KA94}. The shear-stress velocity is almost constant at $z \simeq 0.14$--0.40 m [Fig.~\ref{F2}(c)]. Below and above this constant-stress sublayer, there are the roughness sublayer and the outer sublayer, respectively, where turbulence is affected by the roughness or the outer laminar flow. The logarithmic law that corresponds to the constant stress is unclear in the profile of the mean streamwise velocity [Fig.~\ref{F2}(a)] because our constant-stress sublayer is relatively thin and there is uncertainty in defining the effective origin for the logarithmic law.

\begin{figure}[!]
\resizebox{8cm}{!}{\includegraphics*[3cm,7cm][18cm,27cm]{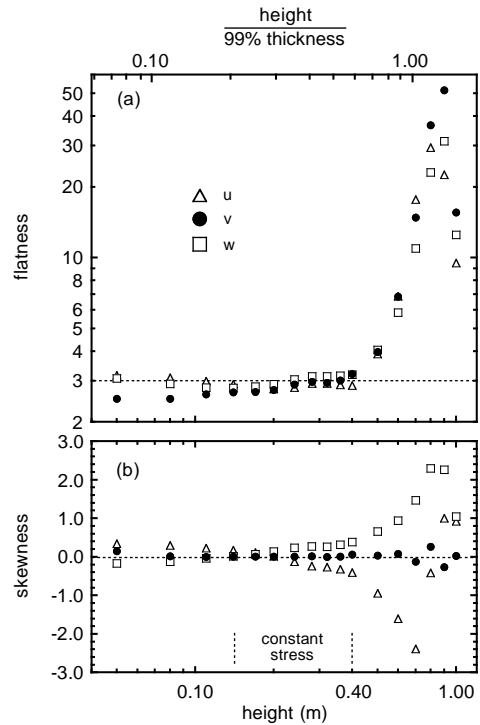}}
\caption{\label{F3} (a) Flatness factors $F_u$, $F_v$, and $F_w$. (b) Skewness factors $S_u$, $S_v$, and $S_w$. The abscissa is the height $z$. The triangles denote the $u$ component, the circles denote the $v$ component, and the squares denote the $w$ component. The horizontal dotted lines indicate the Gaussian values $F=3$ (a) and $S=0$ (b). On the uppermost axis, we indicate the height normalized by the 99\% thickness. We also indicate the $z$ range of the constant-stress sublayer. }
\end{figure}

\section{RESULTS AND DISCUSSION}
\label{S4}

Figure~\ref{F3}(a) shows the flatness factors $F_u$, $F_v$, and $F_w$ as a function of the height $z$. Figure~\ref{F3}(b) shows the skewness factors $S_u = \langle u^3 \rangle / \langle u^2 \rangle ^{3/2}$, $S_v$, and $S_w$. The flatness factors are close to 3 at $z \lesssim 0.40$ m. As the height is increased above $z \simeq 0.40$ m, the flatness factors begin to increase. They have pronounced peaks at $z \simeq 0.90$ m. The skewness factors are also significant at $z \gtrsim 0.40$ m, except for the $v$ component that is free from the shear of the boundary layer. Similar results were obtained in previous works \cite{AJ84,OCI00}.

We focus on the $v$ component, which is best suited to our analysis. The range of flatness factor is widest. The dependence on the height $z$ is simple. At $z \lesssim 0.40$ m, the PDF is sub-Gaussian. At $z \gtrsim 0.40$ m, the PDF is hyper-Gaussian. Only the $v$ component at $z \simeq 0.40$ m exhibits simultaneously the Gaussian values $F=3$ and $S=0$.

\begin{figure}[!]
\resizebox{8cm}{!}{\includegraphics*[3cm,3cm][18cm,27cm]{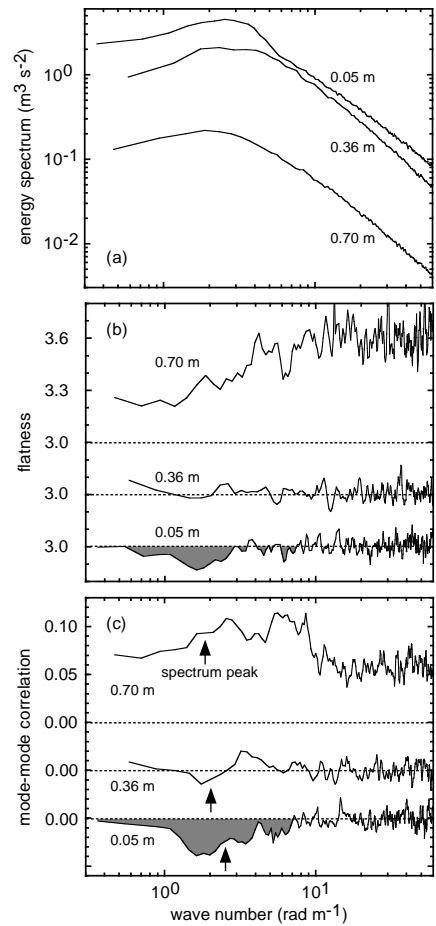}}
\caption{\label{F4} (a) Energy spectrum $E_n$ of the $v$ component at $z = 0.05$, 0.36, and 0.70 m. (b) Flatness factor of the Fourier coefficient $F_n$. (c) Mode-mode correlation $C_{nn_p}$ ($n \neq n_p$). The abscissa is the wave number $k_n$. The horizontal dotted lines indicate the Gaussian value of 3 (b) and the noncorrelation value of 0 (c). The arrows indicate the wave numbers $k_{n_p}$ of the $E_n$ peaks. In (b) and (c), we made moving averages over five adjacent wave numbers. The hatched areas emphasize the sub-Gaussianity and negative correlation of the data at $z = 0.05$ m.}
\end{figure}

Figure~\ref{F4}(a) shows the energy spectrum $E_n$ for the $v$ component at $z = 0.05$, 0.36, and 0.70 m. Figure~\ref{F4}(b) shows the flatness factor of the Fourier coefficient $F_n = \langle a_n^4 \rangle / \langle a_n^2 \rangle^2$. Figure~\ref{F4}(c) shows the mode-mode correlation $C_{nn_p}$ ($n \ne n_p$):
\begin{equation}
\label{eq3}
C_{nn_p} = 
\frac
{\langle a_n^2 a_{n_p}^2 \rangle - \langle a_n^2 \rangle \langle a_{n_p}^2 \rangle}
{(\langle a_n^4 \rangle - \langle a_n^2 \rangle ^2) ^{1/2}
 (\langle a_{n_p}^4 \rangle - \langle a_{n_p}^2 \rangle ^2)^{1/2}},
\end{equation} 
where $n_p$ corresponds to the wave number $k_{n_p}$ of the $E_n$ peak. These quantities were obtained by dividing the data into segments of $2^{15}$ points. We regarded them as independent realizations of turbulence, applied the Fourier transformation individually to them, and calculated averages over them at each of the wave numbers. The energy spectrum was obtained using the Welch window function. Since this and other usual window functions were found to affect significantly the flatness factor or mode-mode correlation, we were forced to obtain them by appending the inverted sequence to each sequence of the segments. Any method to remove effects of discontinuity at the ends of a data segment modifies the data and thereby affects some of the statistics. Our present method is not an exception but happens to have no serious effect on the statistics of our interest.

Velocity fluctuations at $z = 0.36$ m are Gaussian [$F_v = 3.00$; Fig.~\ref{F3}(a)]. Throughout the wave numbers, the flatness factor of the Fourier coefficient is close to Gaussian [Fig.~\ref{F4}(b)], and the mode-mode correlation is absent [Fig.~\ref{F4}(c)]. Thus the Fourier coefficients are Gaussian and independent of each other. The height $z = 0.36$ m is near the upper edge of the constant-stress sublayer and also near the middle of the entire boundary layer. We consider that eddies of various sizes and strengths pass the probe randomly and independently.

Velocity fluctuations at $z = 0.05$ m are sub-Gaussian [$F_v = 2.51$; Fig.~\ref{F3}(a)]. At around the peak of the energy spectrum $E_n$, the flatness factor of the Fourier coefficient is sub-Gaussian [Fig.~\ref{F4}(b)]. This sub-Gaussianity of the Fourier coefficients is associated with that of velocity fluctuations (Sec.~\ref{S2}). Although the former is less significant than the latter, the mode-mode correlation is negative [Fig.~\ref{F4}(c)]. This is also associated with the sub-Gaussianity of velocity fluctuations. Even if the Fourier coefficient at the $E_n$ peak has a large amplitude, its effect to velocity fluctuations tends to be weakened by small amplitudes of the Fourier coefficients at nearby wave numbers, as compared with a noncorrelation case.

The mode-mode correlation shown in Fig.~\ref{F4}(c) is merely a representative example. Similar negative correlations are found for other pairs of Fourier coefficients at around the energy peak. It is possible to have more than two Fourier coefficients that are negatively correlated with each other if the absolute values of the correlation coefficients are small.

The height $z = 0.05$ m is in the roughness sublayer. A plausible explanation is that turbulence is contaminated with bounded-amplitude motions due to wavy wakes of the roughness. The amplitudes are required to be in a bounded range because velocity fluctuations and their Fourier coefficients are sub-Gaussian (Sec.~\ref{S2}). The individual motions are required to contribute to a range of wave numbers, possibly through the presence of spatial structures, because there are mode-mode correlations.

Velocity fluctuations at $z = 0.70$ m are hyper-Gaussian [$F_v = 14.82$; Fig.~\ref{F3}(a)]. Throughout the wave numbers, the flatness factor of the Fourier coefficient is hyper-Gaussian [Fig.~\ref{F4}(b)]. This is associated with the hyper-Gaussianity of velocity fluctuations (Sec.~\ref{S2}). The mode-mode correlation is positive [Fig.~\ref{F4}(c)]. If the Fourier coefficient at the $E_n$ peak has a large amplitude, its effect to velocity fluctuations is strengthened by large amplitudes of the Fourier coefficients at nearby wave numbers.

The height $z = 0.70$ m is near the outer edge of the boundary layer, where turbulence is intermittent \cite{R91,SG00}. There are only eddies that have been ejected from the lower heights. Actually, the skewness factors of the $u$ and $w$ components are negative and positive, respectively [Fig.~\ref{F3}(b)]. The eddies intermittently pass the probe and enhance velocity fluctuations. They are accordingly hyper-Gaussian. The intermittency is also responsible for the hyper-Gaussianity of the Fourier coefficients (Sec.~\ref{S2}). The presence of spatial structures is responsible for the presence of mode-mode correlation. It should be noted that, since turbulence is not space filling, this intermittency is regarded as contamination of turbulence with a laminar flow.

The hyper-Gaussianity of velocity fluctuations in the outer sublayer has been known to be an intermittent phenomenon for a long time, readily understood in the space domain \cite{SG00}. Their Fourier coefficients in the wave number domain are not so useful but have interesting properties as in the cases of sub-Gaussian and Gaussian velocity fluctuations.

\begin{figure}[!]
\resizebox{8cm}{!}{\includegraphics*[3cm,5cm][18cm,27cm]{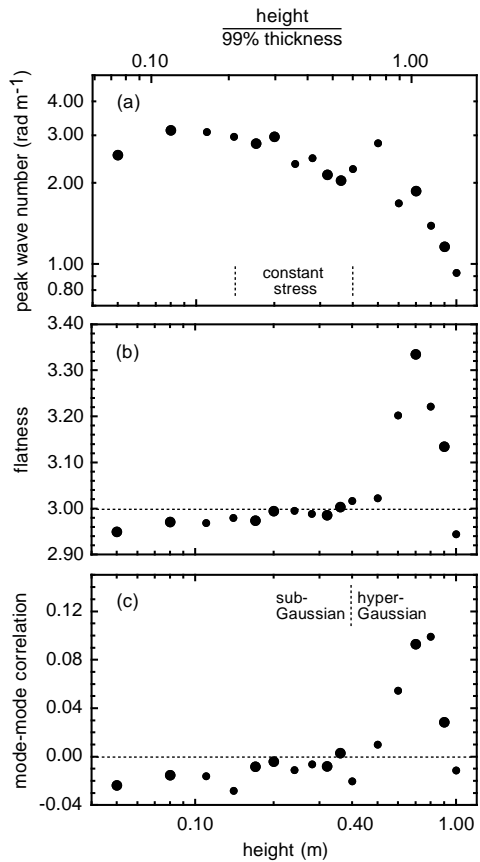}}
\caption{\label{F5} (a) Wave number $k_{n_p}$ of the peak of the energy spectrum of the $v$ component. (b) Flatness factor of the Fourier coefficient $F_n$. (c) Mode-mode correlation $C_{nn_p}$ ($n \neq n_p$). The flatness factor and the mode-mode correlation are medians around $k_{n_p}$ within $2^{\pm 1} k_{n_p}$. The abscissa is the height $z$. The large circles denote data with $32 \times 10^6$ points, while the small circles denote data with $8 \times 10^6$ points. The horizontal dotted lines indicate the Gaussian value of 3 (b) and the noncorrelation value of 0 (c). On the uppermost axis, we indicate the height normalized by the 99\% thickness. We also indicate the $z$ range of the constant-stress sublayer as well as the $z$ ranges where velocity fluctuations are sub- and hyper-Gaussian.}
\end{figure}

Figure~\ref{F5} summarizes behaviors of the Fourier coefficients as a function of the height $z$: (a) the wave number $k_{n_p}$ of the peak of the energy spectrum, (b) the flatness factor of the Fourier coefficient $F_n$, and (c) the mode-mode correlation $C_{nn_p}$ ($n \ne n_p$). The latter two quantities are medians around $k_{n_p}$ within $2^{\pm 1} k_{n_p}$. Since statistical uncertainty is not insignificant, the shorter data with $8 \times 10^6$ points are denoted by the smaller symbol. With an increase of the height, the peak wave number decreases because energy-containing eddies become larger \cite{footnote5}. The behaviors of the flatness factor and mode-mode correlation are in accordance with the behavior of velocity fluctuations. At $z \lesssim 0.40$ m, velocity fluctuations are sub-Gaussian. The Fourier coefficients are sub-Gaussian and exhibit negative mode-mode correlations. At $z \gtrsim 0.40$ m, velocity fluctuations are hyper-Gaussian. The Fourier coefficients are hyper-Gaussian and exhibit positive mode-mode correlations.

\section{CONCLUDING REMARKS}
\label{S5}

The PDF of single-point velocity fluctuations in turbulence is not universal but reflects energy-containing motions. We studied velocity fluctuations using their Fourier coefficients in the energy-containing range. In ideal turbulence where energy-containing motions are random and independent, the Fourier coefficients tend to Gaussian and independent of each other. Velocity fluctuations tend to Gaussian. This is the case at around the middle of a rough-wall boundary layer, where eddies of various sizes and strengths pass the probe randomly and independently. However, if turbulence is contaminated with bounded-amplitude motions such as wavy wakes, the Fourier coefficients tend to sub-Gaussian and their amplitudes are correlated negatively. Velocity fluctuations tend to sub-Gaussian. This is the case in the lower part of a rough-wall boundary layer, where contamination with wavy wakes of the roughness is significant. If turbulence is intermittent or contaminated with a laminar flow, the Fourier coefficients tend to hyper-Gaussian and their amplitudes are correlated positively. Velocity fluctuations tend to hyper-Gaussian. This is the case in the upper part of a rough-wall boundary layer, where turbulence is not space filling and eddies intermittently pass the probe.

We previously studied velocity fluctuations in grid turbulence \cite{M02}. At small distances from the grid, turbulence is developing. There are bounded-amplitude motions due to wavy wakes of the grid. The PDF of velocity fluctuations tends to sub-Gaussian. At intermediate distances, turbulence is fully developed. The PDF tends to Gaussian. At large distances, turbulence is decaying. There remain only strong eddies. They intermittently pass the probe. The PDF tends to hyper-Gaussian. These results for grid turbulence are consistent with our present results for boundary-layer turbulence.

Sreenivasan and Dhruva \cite{SD98} obtained $F_u = 2.66$ in the atmospheric boundary layer. The exact observational condition is unknown to us, but the flow at the measurement position $z = 35$ m could be affected by wavy wakes of the ground roughness. The observed sub-Gaussianity could be attributable to possibly bounded amplitudes of these wakes.

Velocity fluctuations tend to sub-Gaussian in direct numerical simulations of homogeneous, isotropic, stationary turbulence \cite{VM91,J93,GFN02}. Since the simulations were done under forcing over narrow ranges of the smallest wave numbers, the energy spectra are steep and close to the power law $k_n^{\alpha}$ with $\alpha < -1$. The observed sub-Gaussianity could be attributable to the forced motions that dominate the velocity fluctuations \cite{J98,GFN02} (see also Sec.~\ref{S2}). Thus these numerical results are not inconsistent with our experimental results where the energy spectrum is relatively flat in the energy-containing range [Fig.~\ref{F4}(a)]. It is of interest to study such numerical data in the same manner as in the present work.

\begin{acknowledgments}
This research has been supported in part by the Japanese Ministry of Education, Science, and Culture under grant (B2) 14340138. The authors are grateful to the referee for helpful comments.
\end{acknowledgments}


\begin{thebibliography}{999}
\bibitem{footnote2} If turbulence is contaminated with a sinusoidal wave $u(x) \propto \sin (x)$, the correlation length $l_c$ is infinite. It is impossible to arrange the calculation of the Fourier transforms [Eq. (\ref{eq2})] in such a way that the central limit theorem is applicable.
\bibitem{footnote3} Usually, the 99\% thickness and the displacement thickness are defined not with the maximum value $\hat{U}$ of the mean streamwise velocity but with the velocity of the in-coming flow. However, since our wind tunnel was not capable of adjusting its ceiling to set the pressure gradient to be zero, the mean streamwise velocities at the highest positions were greater than the in-coming velocity. We thereby used the maximum value of the mean streamwise velocity obtained at $z = 0.90$ m.
\bibitem{footnote5} For the cases at $z = 0.40$, 0.50, and 0.60 m, we did not use the exact peak wave number, which is unnaturally large possibly due to statistical uncertainties. We instead used the wave number that corresponds to the second greatest value of the energy spectrum. This wave number is 60\%--80\% of the peak wave number.
\end{thebibliography}
\end{document}